\documentclass[12pt]{iopart}
\usepackage{graphicx}

\newcommand{\be}{\begin{equation}}
\newcommand{\ee}{\end{equation}}
\newcommand{\bea}{\begin{eqnarray}}
\newcommand{\eea}{\end{eqnarray}}
\renewcommand\({\left(}
\renewcommand\){\right)}

\newcommand{\gs}{\delta_{\scriptscriptstyle \rm GS}}
\newcommand{\Lkin}{\mathcal{L}_\mathrm{kin}}
\newcommand{\GeV}{\, {\rm GeV}}
\renewcommand{\Re}{\,\mathrm{Re}\,}
\renewcommand{\Im}{\,\mathrm{Im}\,}

\begin{document}

\begin{flushright}
\hfill DESY 07-187\\
\end{flushright}

\title{$D$-term Uplifted Racetrack Inflation}

\author{Ph~Brax$^1$, A~C~Davis$^2$, 
S~C~Davis$^3$, R~Jeannerot$^4$ and M~Postma$^{5,6}$}

\address{ ${}^1$ Service de Physique Th\'eorique, CEA/DSM/SPhT,
Unit\'e de recherche associ\'ee au CNRS, CEA--Saclay 91191 Gif/Yvette
cedex, France}
\address{${}^2$ DAMTP, Centre for Mathematical Sciences,
University of Cambridge, Wilberforce Road, Cambridge,CB3 0WA, UK}
\address{${}^3$ Laboratoire de Physique Theorique d'Orsay,
B\^atiment 210, Universit\'e Paris-Sud 11, 91405 Orsay Cedex, France}
\address{${}^4$ Instituut-Lorentz for Theoretical Physics, Postbus 9506,
 2300 RA Leiden, The Netherlands}
\address{${}^5$ DESY, Notkestra\ss e 85, 22607 Hamburg, Germany}
\address{${}^6$ Nikhef, Kruislaan 409, 1098 SJ Amsterdam, The
Netherlands}

\eads{\mailto{brax@spht.saclay.cea.fr}, 
\mailto{A.C.Davis@damtp.cam.ac.uk}, \mailto{sdavis@lorentz.leidenuniv.nl},
\mailto{jeannero@lorentz.leidenuniv.nl}, \mailto{postma@mail.desy.de}}

\begin{abstract}

It is shown that racetrack inflation can be implemented in a moduli
stabilisation scenario with a supersymmetric uplifting $D$-term. The
resulting model is completely described by an effective supergravity
theory, in contrast to the original racetrack models. We study the
inflationary dynamics and show that the gaugino condensates vary
during inflation.  The resulting spectral index is $n_s \approx 0.95$
as in the original racetrack inflation model. Hence extra fields do not appear
to alter the predictions of the model. An equivalent, simplified
model with just a single field is presented.

\end{abstract}

\section{Introduction}

With the advent of precision cosmological data, the need for a precise
description of inflation is pressing. This is all the more conspicuous
in the light of the forthcoming launch of the Planck
satellite. Unfortunately, inflation model building has produced a
plethora of viable scenarios. Few of them, though, have a sound
physical motivation. Yet it is clear that inflation belongs to the
realm of high energy physics, as it takes place at energies beyond the
TeV scale.  Moreover, a period of inflation cannot be obtained with
just the standard model fields and their interactions.  It may thus be
argued that a tenable determination of the inflation parameters,
leading to predictions for the CMB spectrum, must be achieved within
our models of physics beyond the standard model. A most promising
candidate for this is string theory.

Models of inflation within string theory have been constructed
recently. They all face the closely related moduli-~\cite{kklmmt} and
$\eta$-problem~\cite{eta}. All string moduli, with the exception
of the inflaton fields, need to be fixed during inflation.  The
resulting inflation potential should also be sufficiently flat to
obtain a period of slow-roll inflation. String models of slow roll
inflation fall into two categories. The first one, brane inflation,
uses the motion of branes to achieve inflation~\cite{Dvali}. Warped
throats~\cite{kklmmt} and non-relativistic motion~\cite{dbi1,dbi2} may
be invoked to make the inflaton potential suitably flat.  The second
category is moduli inflation~\cite{modinfl1,modinfl2,race1,race2}.  In this
set-up one of the many string moduli fields plays the role of the
inflaton.  Near an extremum, the potential may be sufficiently flat
for inflation.  The model proposed in this paper is an example of the
latter approach.

The first explicit construction of a string vacuum with all the moduli
stabilised and with a positive cosmological constant is the KKLT
set-up~\cite{kklt}.  In this approach a flux potential stabilises the
dilaton and the complex structure moduli.  The K\"ahler moduli acquire
a non-trivial potential from non-perturbative effects such as gaugino
condensation.  In general, the resulting configuration is a stable AdS
vacuum with a negative cosmological constant and no
supersymmetry-breaking. The lifting to a Minkowski or dS background,
in which SUSY is broken, is achieved with antibranes located at the
tip of the warped throats induced by the internal
fluxes. Unfortunately, these antibranes break supersymmetry explictly.

Within the KKLT framework, and assuming that the non-perturbative
potential is of the racetrack type, it has been shown that saddle
points in the moduli potential exist around which slow roll inflation
can occur~\cite{race1,race2}. One of the successes of these racetrack
inflation models is the value of the spectral index $n_s \approx
0.95$, very close to the WMAP3 results. A drawback of these models is
the explicit breaking of supersymmetry, which means that one must go
beyond the supergravity approximation in which radiative corrections
are under control. In this paper we avoid the problem by building a
racetrack model where the lifting potential is due to a manifestly
supersymmetric $D$-term. The use of uplifting $D$-terms has been
advocated in \cite{bkq,ana}.

Special care must be paid to gauge invariance, anomaly cancellation
and the presence of extra meson fields when constructing the gaugino
condensates. It will be interesting to see how these meson fields can
affect the CMB predictions of the model.  This touches upon another
motivation for our work, which is to test the robustness of the
prediction for the spectral index $n_s \approx 0.95$.  Is this a
generic prediction of inflation with racetrack superpotentials, or
rather a more model dependent prediction?  As we will see, our model also 
gives $n_s \approx 0.95$.  Although this is not a definite answer to the above
question, it does suggest that the prediction is generic. This claim
is supported by the analysis of a three cosine inflation model which
is deduced from the racetrack model by freezing all the fields at
their saddle point values and leaving the imaginary part of the
K\"ahler modulus to roll slowly along the unstable direction. In
particular, the spectral index of the simplified model as a function
of the only free parameter $\eta_{\rm sad}$, the $\eta$ parameter at the
saddle point, gives a very accurate description of the exact result
obtained using the full scalar potential of the racetrack model. In
this sense, the universality of the $n_s\le 0.95$ bound follows from
the corresponding bound obtained for the three cosine model.

The paper is organised as follows. In section~\ref{s:mod} we discuss the
ingredients of the moduli sector, and present our inflation
model. After a short recapitulation of the slow roll formalism in
section~\ref{s:inf}, we give the results of our numerical analysis in
section~\ref{s:num}. We find that they are very similar to those of other
racetrack models. This leads us to propose the simplified effective model of
section~\ref{s:eff}. Our results are summarised in section~\ref{s:con}.

\section{The racetrack model}
\label{s:mod}

In this section we focus on the model building aspects of the $D$-term
uplifted racetrack model of inflation. Numerical results and
comparisons with the original racetrack inflation model can be found
in the following sections. The idea behind the racetrack model of
inflation is that the moduli sector of type IIB string theory
compactified down to 4D plays the role of the inflation sector.  In
the original racetrack inflation papers, the procedure was implemented
within the KKLT scheme whereby the vacuum of the theory is obtained
via a non-supersymmetric uplifting procedure~\cite{race1,race2}.  In
particular, uplifting is achieved using anti-D-branes living at the
tip of a warped throat. One of the drawbacks of these racetrack models
is the absence of an explicit supergravity description for the
uplifting mechanism. In this section we will describe a supersymmetric
uplifting model extending~\cite{ana}.

Let us first describe a minimal supergravity setting wherein the
number of fields has been reduced to only three fields $T$ and
$\Phi_i$ with $i=1,2$. Typically, in type IIB string models, (the real
part of) the modulus $T$ describes the size of the compactification
manifold while the $\Phi_i$ correspond to meson fields on D7 branes.
The dynamics are described by a non-perturbative superpotential.  In
the scenario of \cite{ana}, this potential arises from gaugino
condensation on D7 branes. Racetrack inflation requires a
superpotential with at least two exponential terms, which necessitates
the presence of two non-Abelian gauge groups $SU(N_i)$. Considering
the case of meson fields $\Phi_i$ arising from $N_{fi}$ quark flavours
of each group
\be
W= W_0 + A \frac{\rme^{-a T}}{\Phi_1^{r_1 a}}
 + B \frac{\rme^{-b T}}{\Phi_2^{r_2 b}} \, .
\label{W}
\ee
Without loss of generality, we can take the parameters $W_0,A,B$ to be
real. The constant term arises from integrating out the stabilised
dilaton and the complex structure moduli.  For simplicity we are
considering diagonal quark condensates $\Phi_i^2 = (1/N_{fi})
\sum_{a=1}^{N_f} Q_{ai} \bar Q_{ai}$, where $Q_{ai}$ and $\bar Q_{ai}$
are quark and antiquark fields.

The constants $a,b,r>0$ are given by
\begin{equation}
a=\frac{4\pi k_1}{N_1-N_f} \, , \qquad
b= \frac{4\pi k_2}{N_2-N_f}\, , \qquad
r_i=\frac{N_{fi}}{2\pi k_i}  \, ,
\label{r}
\end{equation}
where  $k_i=\Or(1)$ are parameters relating the gauge coupling
function of the $SU(N_i)$ group to the modulus field $T$. We have
identified
\be f_i = \frac{k_i T}{2\pi} \label{fi} \ee
for any gauge group $G_i$.  The constants in front of the exponents are
related to the gauge parameters by
\begin{equation}
A=(N_1-N_{f1}) (2N_{f1})^{r_1 a/2} \, , \qquad
B=(N_2-N_{f2}) (2N_{f2})^{r_2 b/2} \, .
\label{A}
\end{equation}
The above effective supergravity level description is valid for
$\Re T \gg 1$ corresponding to the weak coupling limit.

The model possesses a pseudo-anomalous gauged $U(1)_x$ symmetry, whose
anomaly is cancelled by the Green-Schwarz mechanism.  This makes use
of the explicit form of the $U(1)_x$ gauge coupling function $f_x =
{k_x T}/{2\pi}$, which is typical for gauge groups located on D7
branes. Gauge invariance implies that the fields transform as
\be
\delta T = \eta^T \epsilon = i \frac{\gs}{2} \epsilon \, , \qquad
\delta \Phi_i = \eta^i \epsilon = -i\frac{\gs}{2r_i}  \Phi_i \epsilon \, ,
\ee
with $\epsilon$ the infinitesimal gauge parameter, and $\gs$ the
Green-Schwarz parameter. With the minimal field content consisting of
quarks and antiquarks with $U(1)_x$ charges $q_{i}$ and $\bar q_{i}$,
the $U(1)_x SU(N_i)^2$ anomaly conditions read
\begin{equation}
\gs = \frac{N_{f1}}{2\pi k_1} ( q_1+\bar q_1)
=\frac{N_{f2}}{2\pi k_2} (q_2+\bar q_2) \, .
\end{equation}
Using the charge assignment of the meson fields $q_{\Phi i}= (q_i+\bar
q_i)/2$ these anomaly conditions are automatically satisfied for the
parameters~\eref{r}. In addition there is the non-trivial $U(1)_x^3$
condition
\begin{equation}
\gs= \frac{1}{3\pi k_x} \left[ N_1 N_{f1}(q_1^3 +\bar q_1^3) + N_2 N_{f2}
(q_2^3+ \bar q_2^3)\right]
\end{equation}
which can be satisfied for suitable parameter choices.

So far, we have only alluded to the necessary stringy ingredients
which lead to our low energy supergravity model.  Although we do not
attempt to provide a complete stringy construction of the model, more
details about the possible embedding in string theory can be
provided. In particular, explicit models may require the existence of
more fields than the ones presented so far in a minimal setting.

First of all, the existence of chiral fields can be obtained by
considering magnetised D7 branes in orientifold models~\cite{dudas}.
The internal magnetic field is related to the effective
Fayet-Iliopoulos term $E/X^3$ [see \eref{VD1} and \eref{VD2}] of the
$U(1)_x$ gauge symmetry. Close to a fixed point and considering 2
stacks of $N_1$ and $N_2$ D7 branes together with a single isolated
brane, we can construct the low energy spectrum which comprises the
quarks $Q_i$ in the $N_i$ representation corresponding to the open
string between the $N_i$ D7 branes and the $U(1)_x$ brane. Antiquarks
are associated to open strings between the stacks of D7 branes and the
orientifold image of the $U(1)_x$ brane (hence $N_{fi}=1$ here). There
is also a field $\zeta$ associated with the open string between
the $U(1)_x$ brane and its orientifold image. We focus on models where
its $U(1)_x$ charge is positive implying that $\zeta$ picks up a
positive mass thanks to the effective Fayet-Iliopoulos term, and is
stabilised at the origin of field space. There are also charged fields
corresponding to the open string between the D7 stacks (and also
between the stacks and their orientifold images).

We focus on $SU(N_{i})$ $D$-flat directions.  Flat directions are
in one to one correspondence with analytic gauge
invariants~\cite{gauge} such as the meson fields $\Phi_{i}$ built
from the quarks and antiquarks. We consider the particular
direction whereby all the analytic gauge invariants vanish except
the mesons $\Phi_{i}$. We specialise even further to the $D$-flat
directions parameterised by $\Phi=\Phi_1=\Phi_2$ (gauge invariance
then requires we set $r_1=r_2=r$). Along this particular
direction, non-perturbative phenomena in the $SU(N_{i})$ gauge
groups lead to the appearance of the gaugino condensation
superpotential we have used. We will not pursue any further the
string construction of the model. In particular, the analysis of
the anomaly cancellation is modified by the presence of $\zeta$
and of the open strings linking the D7 stacks. This is left for
future work.

The full potential of the theory is given by $V=V_F+V_D$ with
\be
V_F = \rme^K \left(K^{I\bar J} D_I W D_{\bar J} \bar W - 3|W|^2\right)
\, , \qquad
V_D =\frac{1}{2 \Re (f_x)}(i \eta^I K_I)^2 \ ,
\ee
where $I, J$ are summed over $T,\Phi$. We will be using the
superpotential~\eref{W}, but with just a single meson field for
simplicity (so $\Phi_i=\Phi$ and $r_i=r$). The last ingredient of the
supergravity model needed to calculate it is the K\"ahler potential
$K$. For gaugino condensation on D7 branes, we have a minimal K\"ahler
for the meson fields, as in \cite{ana}
\be
K = -3\log \left(T + \bar T 
\right) + |\Phi|^2 \, .
\label{K1}
\ee
Defining $T = X + i Y$ and $\Phi=|\phi|$~\footnote{The Goldstone boson
$\mathrm{arg} \, \Phi$ can be gauged away, and becomes the
longitudinal polarisation of the anomalous $U(1)_x$ vector field.  At
the level of the scalar potential it corresponds to a flat direction
in $V$.} the $D$-term is then
\be
V_D = \frac{E}{X^3}\left(1+\frac{2 X \phi^2}{3r}\right)^2 \label{VD1}
\ee
where $E = 9\gs^2 \pi/(16 k_x)$. The $F$-terms give \bea \fl
V_F &=& \frac{\rme^{\phi^2}}{24 X^3} \biggl\{ A^2 \frac{\rme^{-2a
X}}{\phi^{2(1+ar)}}
 \(3 a^2r^2 + 2a[2aX^2 + 6X -3r] \phi^2 +3\phi^4\)
\nonumber \\ \fl && {}
+B^2 \frac{\rme^{-2b X}}{\phi^{2(1+br)}}
 \(3 b^2r^2 + 2b[2bX^2 + 6X -3r] \phi^2 +3\phi^4 + 3 W_0^2 \phi^2\)
\nonumber \\ \fl && {}
+2 A B \frac{\rme^{-(a+b)X}}{\phi^{2+(a+b)r}}
\(3abr^2 + [4abX^2 +6(a+b)X - 3(a+b)r]\phi^2+3\phi^4\) \cos([a-b] Y)
\nonumber \\ \fl && {}
+6 W_0 A  \frac{\rme^{-a X}}{\phi^{ar}} (2aX-ar + \phi^2)\cos(aY)
\nonumber \\ \fl && {}
+6 W_0 B  \frac{\rme^{-b X}}{\phi^{br}} (2bX-br+ \phi^2) \cos(b\,Y)
\; \biggr\} \, .
\label{VF1}
\eea
The kinetic terms obtained from the K\"ahler potential are
\be
\Lkin = \frac{3}{4X^2}\left(\partial_\mu Y \partial^\mu Y
+ \partial_\mu X \partial^\mu X\right)
+ \partial_\mu \phi \partial^\mu \phi \, .
\label{Lkm}
\ee

For completeness, and to analyse the robustness of the model to
changes of the K\"ahler potential, we also consider models with
no-scale K\"ahler potentials
\be
K = -3\log \left(T + \bar T 
-|\Phi|^2/3\right) \, ,
\label{K2}
\ee
in which case the $D$-term is
\be
V_D = \frac{36 E}{X(6 X-\phi^2)^2 }\left(1+\frac{\phi^2}{3r} \right)^2  \, ,
\label{VD2}
\ee
and the $F$-term is
\bea
\fl  V_F &=& \frac{3}{(6X - \phi^2)^2} \times
\nonumber \\ \fl &&
\biggl\{
2 A B \frac{\rme^{-(a+b)X}}{\phi^{2+(a+b)r}}
 \(3 ab r^2 +[2abX + 2abr + 3a + 3b]\phi^2 \) \cos([a-b] Y)
\nonumber \\ \fl && \; {}
+A^2 a \frac{\rme^{-2a X}}{\phi^{2(1+ar)}} \(3 a r^2 +2 [a X+a r +3]\phi^2\)
+6 W_0 A a \frac{\rme^{-a X}}{\phi^{ar}} \cos(a Y)
\nonumber \\ \fl && \; {}
+B^2 b \frac{\rme^{-2b X}}{\phi^{2(1+br)}}  \(3 b r^2 +2 [b X+b r +3]\phi^2\)
+6 W_0 B b \frac{\rme^{-b X}}{\phi^{br}} \cos(b Y)\;
 \biggr\} \, .
\label{VF2}
\eea
With its no-scale form this more closely resembles the potential used
in the original racetrack model. The kinetic terms are
\be
\fl \Lkin = \frac{9}{(6X-\phi^2)^2}\left(3 \partial_\mu Y \partial^\mu Y
+3 \partial_\mu X \partial^\mu X
+ 2 X \partial_\mu \phi \partial^\mu \phi
-2 \phi \partial_\mu \phi \partial^\mu X\right)  \, .
\label{Lkn}
\ee
In the $\phi \to 0$ limit, these two models become identical. We will
see that during inflation the field $\phi$ is small, and that the
inflationary properties of the models are indeed very similar.

\section{Inflation}
\label{s:inf}

With all the ingredients presented in the previous section, we are now
in position to study the structure of inflation in these models. In
the following, the coefficients in the superpotential are free, and
not restricted to the choices \eref{r} and \eref{A}. We will briefly
comment on the compatibility with the gaugino condensation
superpotential as a function of $N_{i}$, $N_{fi}$ and $k_i$.

The potentials \eref{VF1}, \eref{VF2} presented in the previous
section have a very rich structure with many extrema. The minimum is
close to the global SUSY minimum $W_T = 0$ with $X \propto
|a-b|^{-1}$. Hence, $a,b$ are to be close together so that $X \gg 1$
in order to guarantee the validity of the theory.  The minimum can be
lifted and tuned to a Minkowski vacuum with zero energy using the
$D$-term. Inflation takes place near a saddle point of the
potential. A substantial period of inflation giving rise to a nearly
flat spectrum of density perturbations is obtained if the saddle point
is sufficiently flat.  In this context ``sufficiently flat'' means
having slow roll parameters much smaller than unity.  The first slow
roll parameter $\epsilon$, which measures the slope of the potential,
\be
\epsilon = \frac{(\partial_N V)^2}{12 \Lkin V} \, ,
\ee
vanishes at the saddle point.  The second slow roll parameter $\eta$,
which measures the (negative) curvature of the potential, is the
lowest eigenvalue of the matrix
\be
N^a{}_b = \frac{g^{ac} (\partial_c \partial_b V
-\Gamma^e_{cb} \partial_e V)}{V}
\ee
where the metric $g_{ab}$ is given by $\Lkin = (1/2) g_{ab}
\partial_\mu \varphi^a \partial^\mu \varphi^b$,
with $\varphi^a =\{X,Y,\phi\}$.  For generic parameters, $\eta$ will
not be small. The inflaton, i.e.\ the unstable direction at the saddle
point, should be mostly along the $Y$-direction.  Since $Y$ does not
appear in the K\"ahler potential, this will help to solve the
$\eta$-problem. In addition we need to make sure there is no runaway
$X \to \infty$ behaviour from the saddle, corresponding to
decompactifying space, and that the fields do roll towards the
Minkowski minimum.

It is convenient to use the number of $e$-foldings $N =-\ln a$
(normalised so that $N = 0$ at the end of inflation) as a measure of
time. The scales measured by COBE and WMAP leave the horizon $N_*
\approx 55$ $e$-folds before the end of inflation.  Here and in the
following the subscript $*$ denotes the corresponding quantity at COBE
scales.  Slow roll inflation ends when one of the slow roll parameters
becomes greater than one.  In our numerical analysis we use $\epsilon
=1$ to determine the end of inflation.

For the racetrack models under consideration the mass eigenstates of
$N^a_b$ in the orthogonal directions to the inflaton path are large
$m^2/H_*^2 \gg 1$. Therefore it is a very good approximation to ignore
the isocurvature perturbations, leading to effectively single field
inflation.  The scalar power spectrum is then given by
\be
P = \frac{V}{150 \pi^2 \epsilon}
\label{P}
\ee
evaluated $N_*$ $e$-folds before the end of inflation. The COBE
normalisation imposes that $P \approx 4 \times 10^{-10}$.  A second
crucial observable is the spectral index of the inflaton fluctuations:
\be
n_s \approx 1 - \frac{d \ln P}{dN} \approx 1 +2 \eta
\label{ns}
\ee
where for the second equality we used $\epsilon \ll \eta$ for
racetrack models. WMAP3 has measured $n_s =0.95 \pm 0.02$ for a
negligible tensor contribution to the perturbation spectrum \cite{WMAP3}.  
As we will see the tensor contribution is indeed immeasurably small for our
models.

\begin{figure}
\centerline{\includegraphics{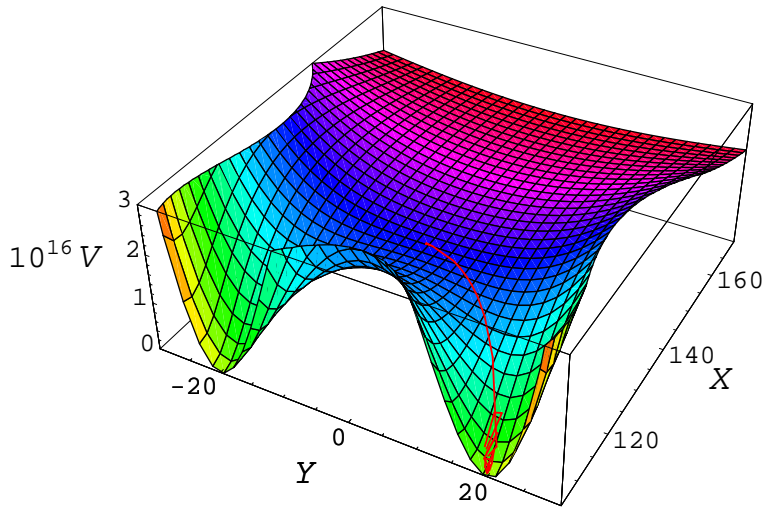}}
\vspace{-1cm}
\caption{Plot of potential $V(X,Y)$, minimised with respect to $\phi$ at
each point. The colour of the shading corresponds the the value of
$\phi$.  The line corresponds to the inflationary trajectory.}
\label{F:VV}
\end{figure}

In general it is not hard to tune the scale of inflation, and get a
power spectrum of the observed size.  The real test for any model is
whether it can produce a scale invariant spectrum as well. This is
also true for racetrack inflation.  Indeed, for the model with a
minimal kinetic term for the meson field~\eref{K1}, the potential has
a scaling property: under a rescaling $L_i \to \lambda L_i$ with $L =
\{a,b,1/r,1/X,1/Y\}$ the potential scales as $V \to \lambda^3 V$. We
can simply choose $\lambda$ to fit the amplitude of the
perturbations~\eref{P} to the COBE spectrum.
Note that if we make the above scaling and scale $\tilde L_i \to
\lambda^{3/2} \tilde L_i$ with $\tilde L =\{W_0,A,B,\sqrt{E}\}$ as
well, the potential remains invariant $V \to V$. If we find one set of
parameters suitable for inflation, it is part of a whole family of
parameter values which all give the same inflationary predictions.
The scaling behaviour of the potential is the same as in the original
racetrack models.  Models with no-scale K\"ahler~\eref{K2} have
similar scaling properties. Combining $L_i \to \lambda L_i$ with
$\phi^2 \to \phi^2/\lambda$, $A\to A/\lambda^{ra}$ and $B \to
B/\lambda^{rb}$, gives $V\to \lambda^3 V$. With the additional scaling
$\tilde L_i \to \lambda^{3/2} \tilde L_i$, the potential will again
remain invariant. Hence for either choice it is straightforward to
adjust the height of the inflationary potential to fit the COBE
normalisation.

In the following, we will describe models where $\eta$ is small and
negative at the saddle points and where the number of $e$-folds is
large enough $N > N_* \approx 55$. To determine the inflationary
trajectory, and the perturbation spectrum, we integrate the equations
of motion numerically, using~\cite{race2}\footnote{There is a sign
difference with
\cite{race2}, as $N$ is defined as the number of $e$-folds
since the beginning of inflation, whereas we set $N$ the number of
$e$-folds before the end of inflation.}
\bea
\frac{d \varphi_i}{d N} &=& -\frac{1}{H} \dot \varphi_i(\pi_i) \nonumber \\
\frac{d \pi_i}{d N} &=&3\pi_i+ \frac{1}{H} (V(\varphi_i)-\Lkin)
\label{eom}
\eea
with $\varphi^a = \{X,Y,\phi\}$ and
$\pi_i = \partial{\Lkin}/{\partial \dot \varphi_i}$. Dots indicate
derivatives with respect to time.

\section{Numerical results}
\label{s:num}


We have analysed the inflaton potential numerically.  As an example we
used the parameters
\be
A =  \frac{1}{20}  \, , \qquad
B = -\frac{9}{100} \, , \qquad
a = \frac{\pi}{50} \, , \qquad
b = \frac{\pi}{45} \, ,
\label{par1}
\ee
for various choices of $r$, $W_0$ and the two K\"ahler potentials.
The parameter $E$ is determined numerically by requiring that the
minimum of the potential is a Minkowski vacuum. The saddle points are
located at $Y_{\rm sad}= 0$. The results are shown in table~\ref{tab}
for the values of $W_0$ which maximise the spectral index $n_s$. In
each case the maximum value was $n_s \approx 0.948$, which is the same
as that obtained in both the original racetrack
papers~\cite{race1,race2}.

\begin{table}
\begin{tabular}{|c@{\, \,}c|cc|ccc|cc|c|}
\hline
K\"ahler & $2 \pi r$  &  $10^{10} E$ & $10^5 W_0$
&  $X_{\rm min}$ & $Y_{\rm min}$ & $\phi_{\rm min}$
&  $X_{\rm sad}$ & $\phi_{\rm sad}$ & $\eta_{\rm sad}$ \\
\hline
Minimal & 1 & 9.9 & -6.5268 &
105.03 & 19.96 & 0.00615 &  127.28 & 0.0032 & -0.0095\\
 & 100 & 13.8 & -9.8224 &
135.28 & 19.22 & 0.153 & 164.87 & 0.0768 & -0.0004 \\
 & 263 & 13.6 & -11.559 &
166.30 & 16.63 & 0.251 &  201.14 & 0.152 & -0.0081 \\
\hline
No-scale & 1  &  9.8 & -6.4974 &
104.64 & 19.93 & 0.0891  &  126.83 & 0.0507 & -0.0102 \\
 & 68  & 4.7 & -5.898 &
102.38 & 15.85 & 1.925 & 120.09 & 1.367 & -0.0095 \\
\hline
\end{tabular}
\caption{Minima and saddle points for parameters~\eref{par1},
with various $r$ and K\"ahlers.}
\label{tab}
\end{table}

\begin{figure}
\centerline{
\includegraphics[width=8cm]{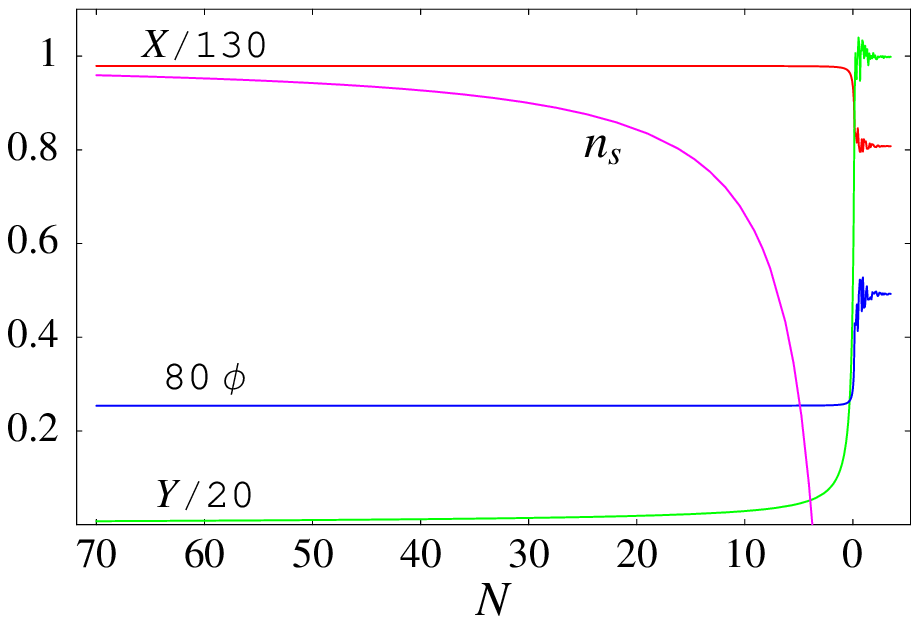}
\hspace{-0.6cm}
\includegraphics[width=8cm]{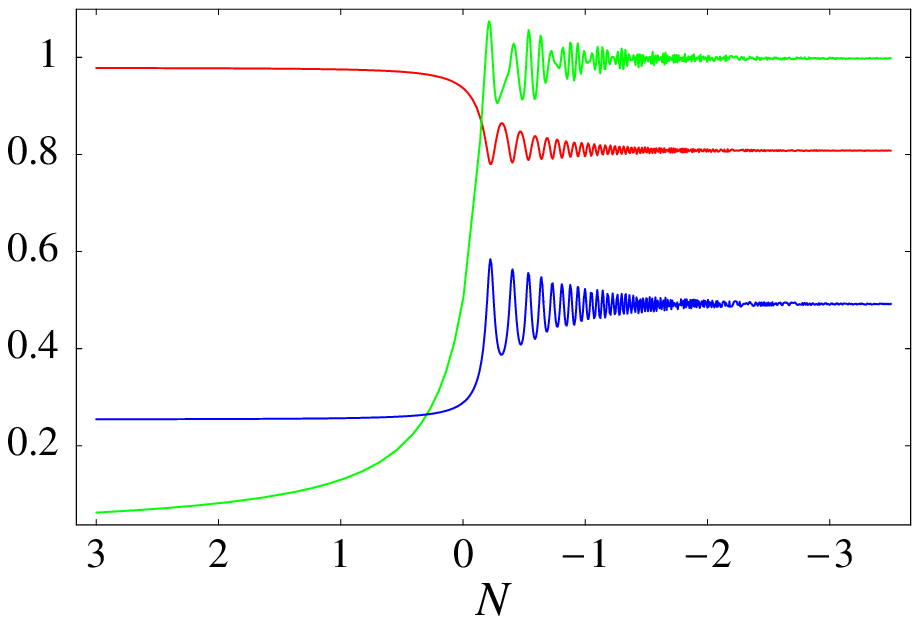}}
\caption{Evolution of the (rescaled) fields as function of the number of
$e$-folds $N$ before the end of inflation.  COBE scales leave the
horizon at $N_* \sim 55$.  Left plot also shows the spectral index
$n_s$.  This and all other plots for parameters~\eref{par1} and the top line
of table~\ref{tab}.}
\label{F:flds}
\end{figure}

Let us take a closer look at the first entry of table~\ref{tab}, which
corresponds to a minimal K\"ahler potential for the meson field, and
has $2\pi r=1$.  The corresponding potential is shown in
figure~\ref{F:VV}.  We integrated the equations of motion~\eref{eom}
numerically for this system, using initial conditions where the field
starts at rest close enough to the saddle so that inflation lasts at
least $N > 55$ $e$-folds. The results are shown in
figure~\ref{F:flds}. As anticipated, initially the inflaton direction
is predominantly along the $Y$-direction.  Indeed, at the time COBE
scales leave the horizon, the $X$ and $\phi$ fields are practically
still at their values at the saddle, whereas $Y$ has moved slightly.
It is only towards the end of inflation, and during the oscillations
after inflation as the fields approach their minimum, that all three
fields are evolving.  This can be appreciated by comparing the field
values at the various points of interest
\bea
X_{\rm sad} = 127.28, \qquad
&Y_{\rm sad} = 0, \qquad
&\phi_{\rm sad} = 0.00317
\nonumber \\
X_* = 127.27, \qquad
&Y_* = 0.18, \qquad
&\phi_* = 0.00317
\nonumber \\
X_{\rm end} = 121.79, \qquad
&Y_{\rm end} = 10.00, \qquad
&\phi_{\rm end} = 0.00360
\nonumber \\
X_{\rm min} =105.03, \qquad
&Y_{\rm min} = 19.96, \qquad
&\phi_{\rm min} = 0.00615
\eea
%
After inflation the scale of SUSY breaking and all soft mass scales
are set by the gravitino mass $m_{3/2} \approx 7 \times 10^7 \GeV$.

The amplitude and spectral index for the above parameters are $P=3.4
\times 10^{-10}$ and $n_s \approx 0.948$. The parameter $W_0$ has been
chosen to maximise the spectral index, as illustrated in
figure~\ref{F:ns}a.  Although the potential can be tuned to be
arbitrarily flat at the saddle $\eta \to 0$, we nevertheless have not
been able to obtain $n_s \approx 1 + 2 \eta$ at COBE scales larger
than $0.948$.  One can argue, as in \cite{race1}, that since the
parameter points which gives the largest $n_s$ also give the most
$e$-folds of inflation, we are most likely to live in a part of the
universe with the maximum spectral index.  This then leads to the
prediction $n_s \approx 0.95$. Figure~\ref{F:flds} shows $n_s$ as a
function of the number of $e$-folds $N$. Around the COBE scale the
$N$-dependence is small. The running spectral index $d n_s/d \ln k
\approx - d n_s/ d N = -5 \times 10^{-3}$ is unmeasurably small. The
same goes for the tensor contribution, since the Hubble parameter
during inflation is $H_* \approx 2 \times 10^{10} \GeV$, which is far below the
measurable values of order $10^{14} \GeV$.

Finally, we would like to comment on the string origin of the
parameters in our $D$-term uplifted racetrack model.  Just as in the
original racetrack model, validity of the supergravity theory $X \gg
1$ is guaranteed for $|a -b| < 1$.  In terms of the gauge parameters
this implies unusually large gauge groups, with $|N_1 - N_2| \ll N_1$.
However, the main obstacle to embedding our model in string theory is
the low value of the Hubble constant during inflation (fixed to fit
the COBE data), and consequently the low gravitino mass in the vacuum
after inflation. This requires taking $E \sim 10^{-9}$ in \eref{VD1},
which translates into a Green-Schwarz parameter $\gs \sim
10^{-5}\sqrt{k_x}$. It is not clear how to get such a low
value~\cite{nilles,dudas,warping}.  A solution to this problem could
be to add an extra field to the dynamics, whose purpose is to
partially cancel the $D$-term as in
\cite{Lalak}. Such a cancelling field could be identified with the field
$\zeta$ between orientifold images of $U(1)$ branes. This is under
investigation.

\section{An effective racetrack model}
\label{s:eff}

In the $r\to0$ limit (and $V$ minimised by $\phi = 0$) our model
approaches the original racetrack model with just one modulus field
$T$. It is thus not surprising that even for $r \neq 0$, but $\phi \ll
1$, our model shares many features of the original model. In both
models $T$ is minimised close to the $W_T = 0$ global SUSY minimum. $X
\gg 1$ is needed to guarantee the validity of the theory,  which is
assured for $|a-b| \ll a,b$.  The presence of a minimum
and a saddle at $Y=0$ puts additional constraints on the parameters:
$A+B <0$, $W_0 <0$ and $a<b$.  Near the saddle the inflaton is mostly
along the $Y$-direction, and isocurvature perturbations are
negligible.

However, quantitatively the models differ by $\Or(1)$ factors. In
our example~\eref{par1} with $2\pi r=1$ we choose the same
$\{a,b\}$ values as in the example discussed in the original
racetrack paper. Our values for $W_0$, $H_*$, and $X$, $Y$ at the
minimum are of the same order of magnitude, but are {\it not} the
same. It is therefore very surprising, that despite the $\Or(1)$
quantitative differences in parameter and field values, we also
found $n_s \leq 0.95$, irrespective of the exact form of the
K\"ahler potential for the $\phi$-field. The improved racetrack
model~\cite{race2}, which employs two moduli fields $T_i$, also gives $n_s \leq
0.95$.  This therefore seems to be a robust prediction of inflation with
racetrack potentials.

\begin{figure}
\centerline{\includegraphics[width=8cm]{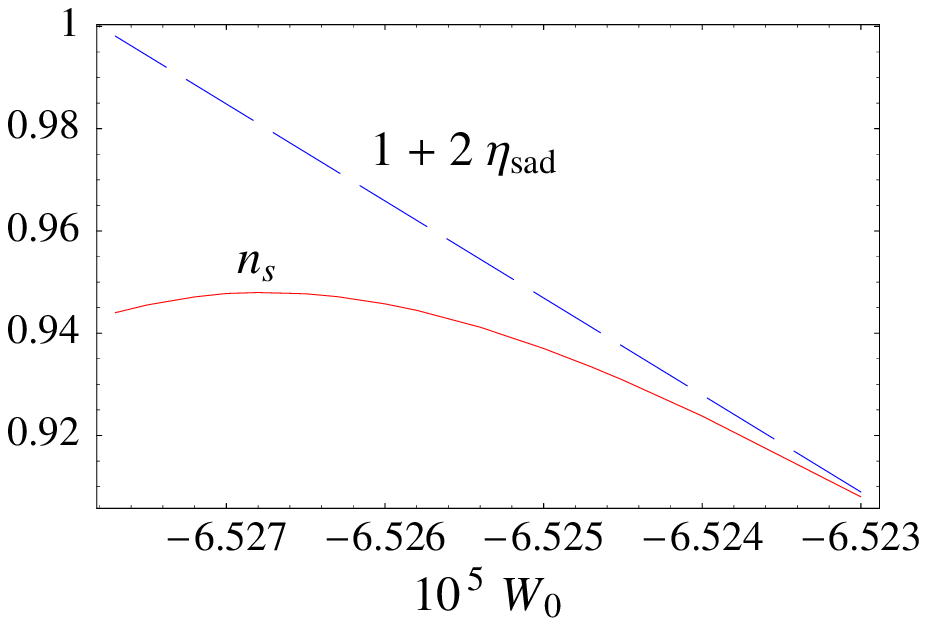}
\hspace{-2cm} (a) \hspace{0.7cm}
\includegraphics[width=8cm]{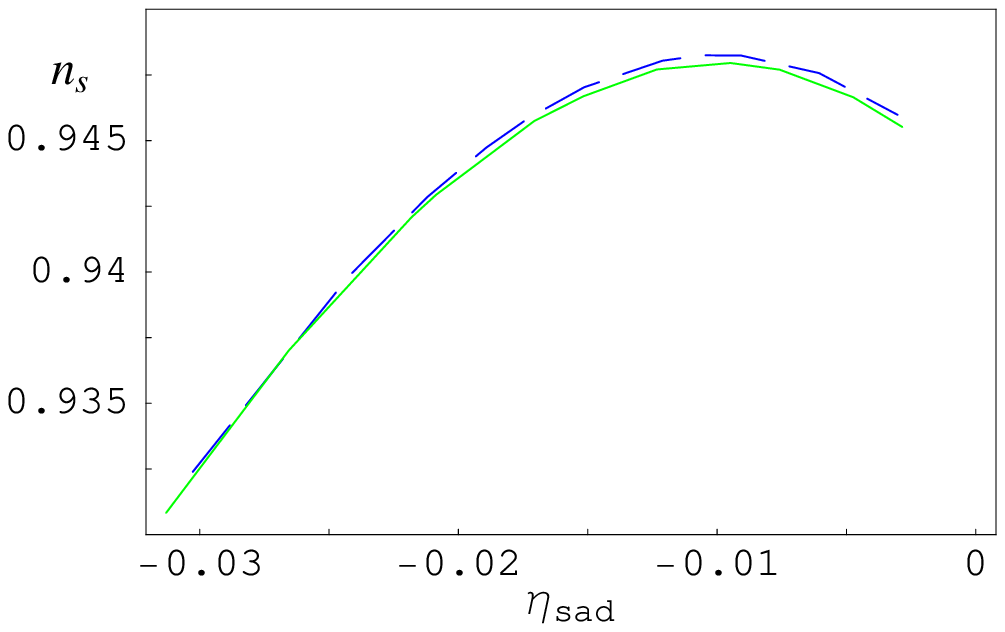} \hspace{-1.5cm} (b) \hspace{1cm}}
\caption{ (a) Variation of spectral index $n_s$ and the initial
value of slow-roll parameter $\eta_{\rm sad}$ as a function of
$W_0$. (b) The spectral index $n_s$ as a function of the value of
$\eta_{\rm sad}$ at the saddle for the racetrack model (solid
green) and the simplified potential~\eref{Vy} (dashed blue).  The
parameters are chosen as in \eref{par1}.} \label{F:ns}
\end{figure}

In all three versions of the racetrack model the potential during
inflation when COBE scales leave the horizon, is of the form
\be
V = V_0 + V_1 \cos(aY) + V_2 \cos(b Y) + V_3 \cos([a-b]Y)
\label{Vy}
\ee
with $Y$ the slowly rolling inflaton field, and the other fields
nearly constant. Indeed in our model $X,\phi$ are nearly fixed near
the saddle, and one can approximate the $X,\phi$-dependent functions
$V_i$ with constants.  In this approximation the kinetic term $\Lkin =
3 (\partial Y)^2 /(4X^2)$ can be made canonical by a constant
rescaling of the field $Y$.  The only difference with the racetrack
models is that the $V_i$ and the kinetic term change {\it after}
observable inflation, when the ``spectator'' fields $X,\phi$ relax to
their values at the minimum.  It will be interesting to see whether
the presence of these spectator fields can affect the inflationary
predictions. If, on the other hand, all physics were to be well captured by the
potential~\eref{Vy} above, then this would explain the universality of
racetrack model predictions for the inflationary observables.

The potential~\eref{Vy} is sufficiently simple that we can explicitly scan the
parameter space for saddle points which are flat enough for inflation,
at least numerically\footnote{Here we restrict ourselves to
potentials with saddles at $Y=0$, as has been done in all analyses of
the various racetrack models.}. To this end we solve for the following
conditions:
\begin{enumerate}
\item  there is a minimum at $Y=Y_0$: $V'(Y_0) =0$ (and $V''(Y_0) > 0$)
\item  the cosmological constant is zero: $V(Y_0) =0$
\item  the saddle at $Y=0$ is flat: $V''(0) \approx 0$
\label{cond3}
\item  the potential is scaled $V \to \lambda V$ to fit the observed power spectrum: $P = 4 \times 10^{-10}$.
\end{enumerate}
These four conditions fix all the $V_i$. The values of $a$ and $b$ are chosen
as in~\eref{par1}.  The resulting spectral index depends only on the value of
$\eta$ at the saddle; for different values for $X,Y_0,a,b$, but the
same $\eta_{\rm sad}$ (determined by the precision to which condition
(\ref{cond3}) above is solved) the spectral index is the same. The dependence
on $\eta_{\rm sad}$ is small. This is shown in figure~\ref{F:ns}b, where
$n_s$ is plotted as a function of $\eta_{\rm sad}$. For all
$\eta_{\rm sad} < -0.02$, guaranteeing enough $e$-folds of inflation, the
spectral index $n_s \approx 0.94 - 0.948$.

Figure~\ref{F:ns}b also shows the corresponding results for our $D$-term
uplifted racetrack model.  The prediction for the spectral index is
practically identical. We can thus conclude that the physics of racetrack
is well captured by the simpler potential~\eref{Vy}.  This explains
the universal prediction for the spectral index $n_s \approx 0.948$ in
all racetrack models.

Although the spectral index only depends on $\eta_{\rm sad}$, and not
the other parameters, the degree of fine tuning of the potential does
depend on those parameters. For example, taking $Y_0 =20$, $X =100$
and $a,b$ as in \eref{par1}, we find that tuning $\eta_{\rm sad}$ at
the 1\% level requires tuning $V_i$ at the level of
$10^{-6}$~\footnote{I.e.\ changing the parameters $\delta V_i \sim
10^{-6} V_i$ will give $\delta \eta_{\rm sad} \sim 0.01$. This agrees
with the tuning in the racetrack scenario: changing $\delta W_0 \sim
10^{-6} W_0$ will change $\delta \eta_{\rm sad} \sim 0.01$ for the
parameters~\eref{par1} and $2\pi r = 1$ (the first entry of
table~\ref{tab}).}.  Increasing $X$ by an arbitrary factor decreases
this tuning by about the same factor. The same applies for a scaling
of $Y_0$, or a simultaneous scaling of $a,b$.  The degree of fine
tuning is most sensitive to $|a-b|$, and the larger their difference
the smaller the tuning.  For example, taking $\pi b = \{49,45,35\}$
changes the level of fine tuning to $\{5\times 10^{-10},8\times
10^{-7}, 1 \times 10^{-6} \}$.  We note however, that in a specific
racetrack model, the values of $X,a,b$ are related, and cannot be
adjusted independently.  Moreover $|a-b|$ needs to be sufficiently
small to assure $X \gg 1$.

\section{Conclusion}
\label{s:con}

In this paper we have presented a racetrack model embedded in a
moduli stabilisation scenario with an uplifting $D$-term. In
addition to the volume modulus $T$, a meson field is
added to the model to guarantee gauge invariance. The resulting
model is completely described by an effective supergravity theory,
in contrast to the original racetrack models.

The model is qualitatively similar to the other inflation models based
on a racetrack superpotential.  For certain parameter choices the
volume modulus is stabilised in a Minkowski vacuum with $\Re T \gg 1$,
ensuring the validity of the effective theory.  Inflation takes place
near a saddle of the potential.  During the initial stages of
inflation, including the time observable scales leave the horizon, the
slowly rolling inflaton field can, to a very good approximation, be
identified with $\Im T$. However during the final stages of inflation
all fields come into play.

The height of the potential can be scaled to fit the COBE
normalisation of the density perturbations.  Tuning is needed to
make the saddle sufficiently flat for inflation and to get a
spectral index that is slightly red shifted.  Although the saddle
point can be tuned arbitrarily flat, we found an upper bound on
the spectral index $n_s \leq 0.95$. This result for the index is
the same as in the original racetrack models, and seems to be a robust
feature of racetrack inflation. This follows from the approximate
equivalence of racetrack inflation to a single field model, obtained by
truncation of the full dynamics, in which only $\Im T$ is allowed vary.

\ack
For financial support, SCD and RJ thank the Netherlands Organisation
for Scientific Research (NWO), and ACD thanks PPARC for partial support.
PhB acknowledges support from RTN European programme MRN-CT-2004-503369.

\end{document}